\documentclass{aastex631}

\usepackage{color}
\usepackage{natbib}
\usepackage{amsmath}

\newcommand{\expf}[1]{{\rm e}^{#1}}
\newcommand{\pot}[2]{{{#1}\times10^{#2}}}

\newcommand{\authourlist}
{
\author[0000-0002-0463-6394]{Gilbert P. Holder}
\affiliation{Astronomy Department, University of Illinois at Urbana-Champaign, 1002 W. Green Street, Urbana, IL 61801, USA}
\affiliation{Department of Physics, University of Illinois Urbana-Champaign, 1110 W. Green Street, Urbana, IL 61801, USA}

\author[0000-0003-3725-6096]{Jens Chluba}
\affiliation{Jodrell Bank Centre for Astrophysics,
The University of Manchester, Manchester, M13 9PL, U.K.}
}

\begin{document}

\title{The radio SZ effect as a probe of the cosmological radio background}

\authourlist

\begin{abstract}
    If there is a substantial cosmological radio background, there should be a radio Sunyaev-Zeldovich (SZ) effect that goes along with it. The radio background Comptonization leads to a slight photon excess at all wavelengths, while Comptonization of the CMB at low frequencies leads to a decrement. For levels of the radio background consistent with observations, these effects cancel each other around $\nu\simeq 735$~MHz, with an excess at lower frequencies and a decrement at higher frequencies. 
    Assuming a purely cosmological origin of the observed ARCADE radio excess, at $\nu \lesssim 20\,{\rm GHz}$ the signal scales as $\Delta T / T_{\rm CMB}\simeq 2\,y\left[ (\nu/735\,{\rm MHz})^{-2.59}-1\right]$ with frequency and the Compton-$y$ parameter of the cluster.
    For a typical cluster, the total radio SZ signal is at the level of $\Delta T\simeq 1\,{\rm mK}$ around the null, with a steep scaling towards radio frequencies. This is above current raw sensitivity limits for many radio facilities at these wavelengths, providing a unique way to confirm the cosmological origin of the ARCADE excess and probe its properties (e.g., redshift dependence and isotropy). We also give an expression to compute the radio-analogue of the kinematic SZ effect, highlighting that this might provide a new tool to probe large-scale velocity fields and the cosmic evolution of the radio background.
\end{abstract}

\section{Introduction}
The thermal Sunyaev-Zeldovich (SZ) effect \citep{Zeldovich1969} is the Comptonization of the relatively cold (2.7~K) cosmic microwave background (CMB) by hot electrons in collapsed halos ($\gtrsim 10^7$~K in galaxy clusters). It is now well-measured in thousands of galaxy
clusters \citep{Planck2013SZ, Hilton2018, bleem2015ApJS}, making them a unique tool for cosmology \citep[e.g.,][]{Carlstrom2002, Mroczkowski2019}.

The SZ effect traditionally refers to the Comptonization of the CMB, but all photons that traverse a galaxy cluster have a probability of scattering and exchanging energy with the cluster gas, including the light from the cosmic infrared background or the cosmic radio background. This was noted by \citet{Cooray2006} as a way to probe the very low-frequency radio background expected from 21cm absorption or excess emission at the end of the dark ages.

In recent years, evidence has emerged that there is a substantial radio background at low frequencies, becoming more important than the CMB at frequencies somewhat below 1 GHz \citep{Fixsen2011excess, Seiffert2011, Singal2018, Dowell2018}. Just as the SZ effect shifts the energies of CMB photons, there should be a similar systematic shift of the radio background spectrum toward higher energies if this background is indeed cosmological. Known radio source counts lead to an inevitable baseline radio background for scattering in the local universe. 

In this work, we calculate the magnitude of the shift expected for the radio background, and show that it can be larger than the traditional SZ effect (and with an opposite sign) for frequencies below $\simeq$735\,MHz. This is a characteristic signal that will depend only on the (often well-measured) Compton $y$-parameter of the galaxy cluster and the amplitude of the radio background. Measuring this signal will confirm the existence of a bright radio background, and doing this as a function of redshift could allow a probe of the redshift evolution of this background. Similarly, by using large samples of clusters, the isotropy of the signal can be probed. 
The equivalent of the kinematic SZ effect \citep{Sunyaev1980} from the scattering of the radio background is also discussed below; it can be used to constrain the large-scale velocity field and as an independent probe of the radio background. Importantly, the spectral degeneracy of the standard kinematic SZ effect with the primordial CMB temperature fluctuations could be avoided and hence may deliver more robust constraints.

\newpage
\section{Comptonization of the Radio Background}
The radio synchrotron background has been measured to be\footnote{This is the Rayleigh-Jeans brightness temperature. This is given by $T=\frac{c^2 }{2k\nu^2}\,I_{\nu}$ in terms of the intensity and $T= \frac{h \nu }{k}\,n_\nu$ in terms of the photon occupation number, $n_\nu$.} \citep{Fixsen2011excess}
\begin{equation}
\label{eq:Tb_AC}
    T_{\rm R, 0}(\nu)= 
    T_{\rm R}(z=0, \nu)= 
    (24.1 \pm 2.1)\,{\rm K} \left[\frac{\nu}{\nu_0}\right]^{-2.59\pm 0.04},
\end{equation}
where $\nu_0=310$~MHz. This is substantially larger than what can be explained by the observed counts of radio-emitting galaxies \citep{condon2012}. This excess is consistent across several independent measurements, and its origin remains unknown \citep[see][for discussion of some non-standard explanations]{Chluba2015GreensII, Trombetti2019, Bolliet2020PI}. In what follows we assume that this is indeed a cosmic radio background that is of extragalactic origin, consistent with the non-detection of bright radio halos around galaxies similar to ours \citep{singal2015}. Its existence at high redshift would have important implications for searches for redshifted 21cm emission from the epoch of the first stars \citep{Chang2018}.
Assuming that this radio background is truly cosmological and isotropic, this same background should be present in galaxy clusters.

For Comptonization of radiation in a galaxy cluster, the important parameter is the
Compton $y$-parameter, defined as $y\equiv \int \frac{kT_{\rm e}}{m_{\rm e} c^2} \,{\rm d}\tau$, where $\tau = \int {\rm d}\tau$ is the line-of-sight optical depth to Thomson scattering in the galaxy cluster and $T_{\rm e}$ is the electron temperature.  For massive galaxy clusters it is possible to get $y\simeq 10^{-4}$. 
To estimate the effect, the Kompaneets equation \citep{Kompa56} can be applied:
\begin{align}
\label{eq:Kompaneets}
\frac{\partial n}{\partial y}
&=\frac{1}{x^2}\frac{\partial}{\partial x}\left( x^4 \left[\frac{\partial n}{\partial x}  + \frac{T_z}{T_{\rm e}} n(1+n) \right] \right),
\end{align}
where $n$ denotes the (isotropic) photon occupation number, $x=h\nu / k T_z$ with $T_z=T_{\rm CMB}(1+z)$ and $T_{\rm CMB}=2.7255\,{\rm K}$. In galaxy clusters, where $T_z\ll T_{\rm e}$, the last terms in Eq.~\eqref{eq:Kompaneets},  describing stimulated scattering effects, can be neglected for practical purposes, yielding the simpler expression \citep[see also][]{Zeldovich1969}
\begin{align}
\label{eq:scattering}
\frac{\partial n}{\partial y}
&\approx\frac{1}{x^2}\frac{\partial}{\partial x} x^4 \frac{\partial n}{\partial x}.
\end{align}
Inserting the standard CMB blackbody occupation, $n_{\rm bb}=1/(\expf{x}-1)$, one finds the well-known SZ formula for the change in the CMB intensity in the direction of a cluster
\begin{align}
\left.\frac{\Delta I}{I}\right|_{\rm tSZ}
&\approx y \frac{x\,\expf{x}}{(\expf{x}-1)}\left[x \coth\left(\frac{x}{2}\right)-4\right].
\end{align}
Equation~\eqref{eq:Tb_AC} implies an additional contribution to the photon occupation number of 
\begin{align}
\Delta n_R(x, z)\approx \frac{T_{\rm R}(\nu, z)}{T_z\,x}\approx
\frac{(8.84 \pm 0.77)\,f(z)}{x} \left[\frac{x}{x_0}\right]^{-2.59\pm 0.04}
=
\frac{(8.84 \pm 0.77)\,f(z)}{x_0} \left[\frac{x}{x_0}\right]^{-3.59\pm 0.04},
\end{align}
with $x_0=h\nu/kT_{\rm CMB}\approx \pot{5.46}{-3}$.
Here, we assumed the redshift-dependence $T_{\rm R}(\nu, z)=T_{\rm R, 0}(\nu)\,f(z)(1+z)$. The astrophysical form factor $f(z)$ represents the fraction of the radio background that is already in place by redshift $z$. One would expect $f(z)=1$ if the radio excess is truly cosmological and fully formed at redshifts much higher than that of the cluster. Known source counts provide a minimum $f(z=0)\simeq 0.2$ \citep{Seiffert2011} that will evolve substantially with redshift and could also evolve with frequency. 

\begin{figure}
\centering 
\includegraphics[angle=0,width=0.63\columnwidth]{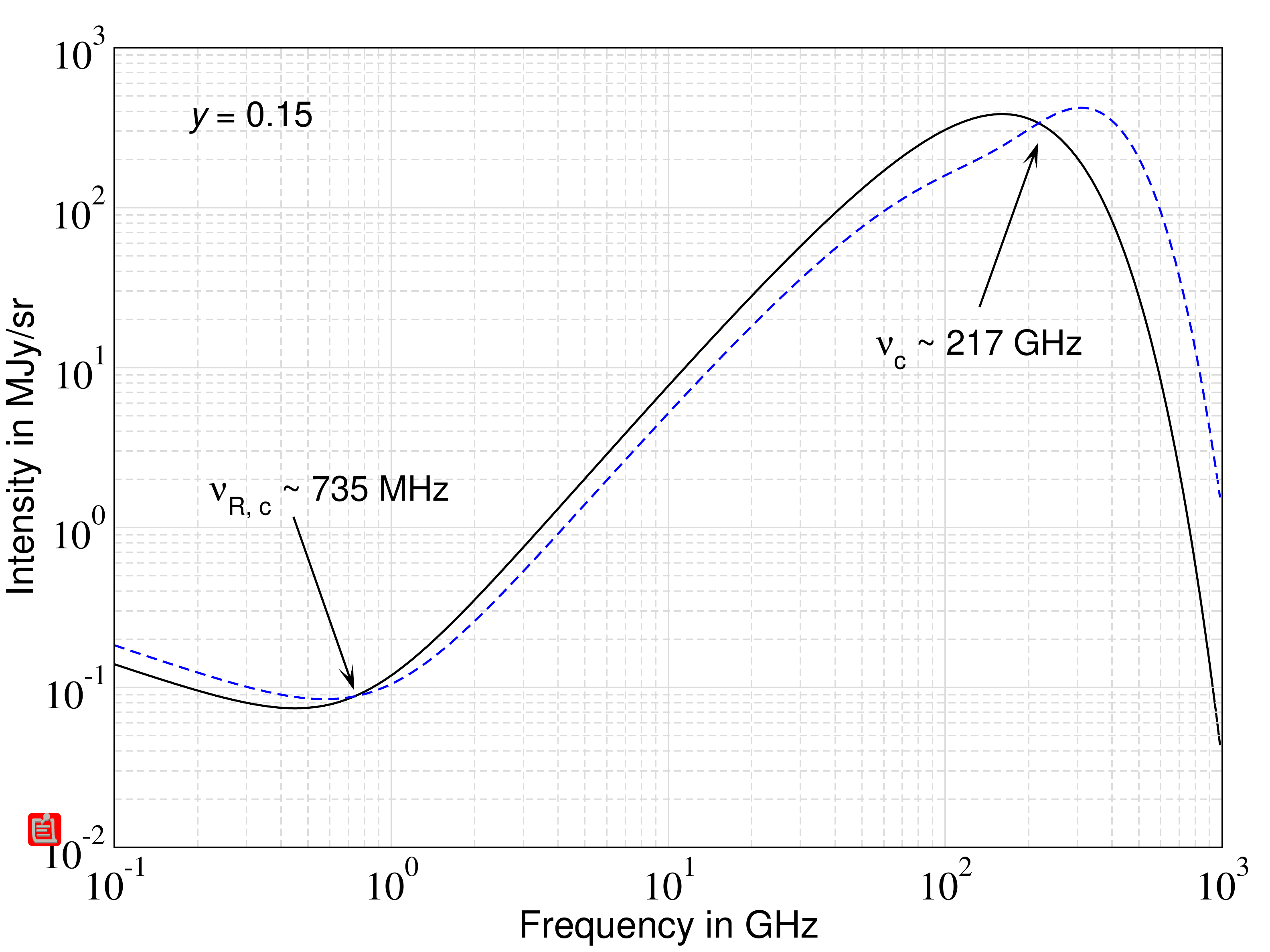}
\\
\caption{Illustration for the total SZ scattering effect. For this we chose an extremely large $y$-parameter of $y=0.15$, applying the single-scattering expression. Typical SZ clusters have $y\simeq 10^{-4}$ implying that the signal is significantly smaller. 
The solid black line shows the unscattered CMB + cosmic radio background, while the dashed blue line includes the effect of Compton scattering inside the galaxy cluster. The well-known tSZ null is visible at high frequencies, while the novel radio SZ null can be appreciated at low frequencies. With respect to the unscattered background, clusters appear as sources at $\nu \lesssim 735\,{\rm MHz}$ and  $\nu \gtrsim 217\,{\rm GHz}$, while they are 'shadows' in between.}
\label{fig:Intensity}
\end{figure}

\begin{figure}
\centering 
\includegraphics[angle=0,width=0.63\columnwidth]{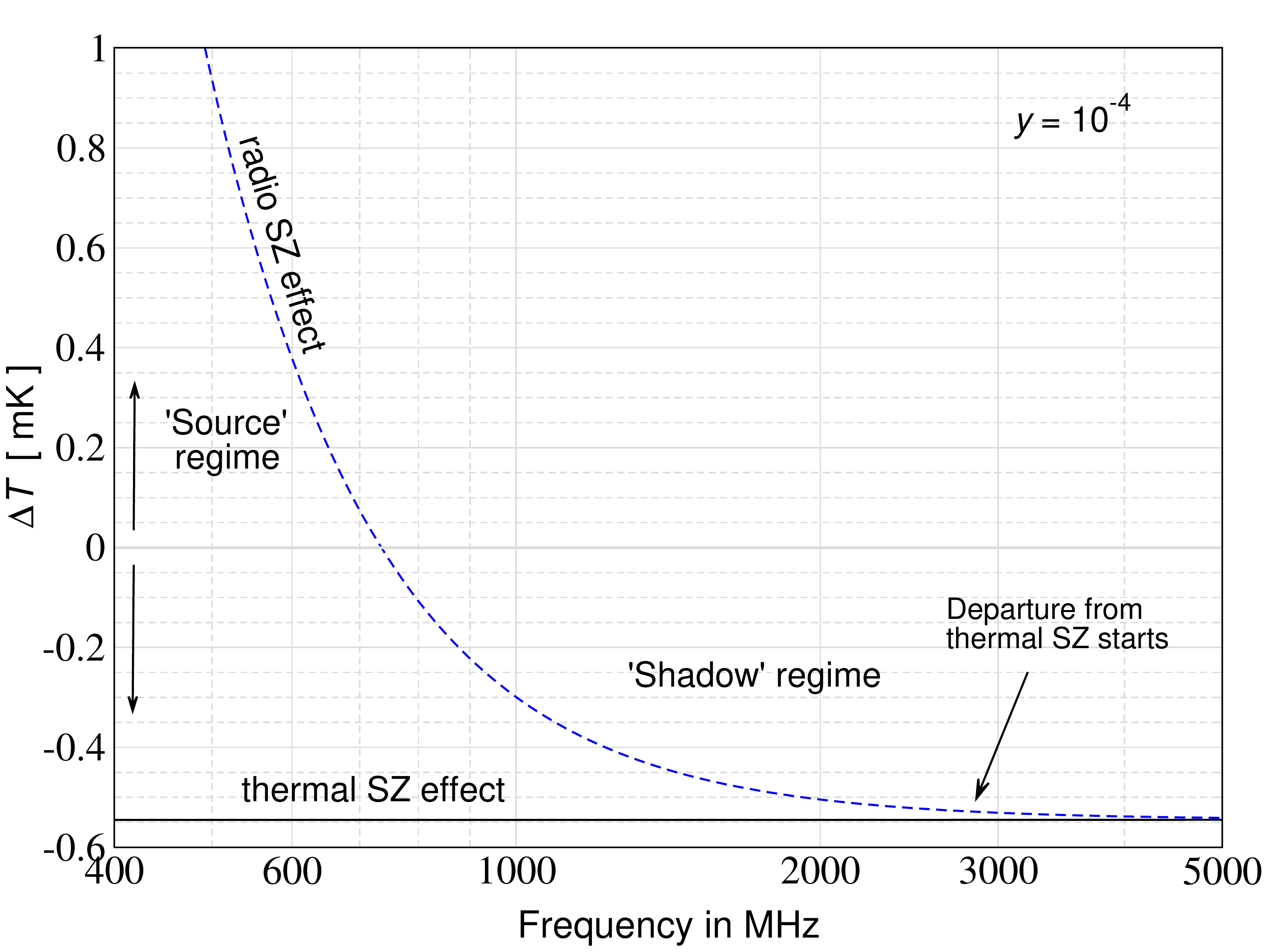}
\\
\caption{Change of the brightness temperature in the direction of a cluster with $y=10^{-4}$ due to the standard thermal SZ effect (solid black line) and the combined total radio SZ effect (dashed blue line). For the standard SZ effect, the cluster will always appear as a 'shadow' in the radio bands. The up-scattering of photons from the radio excess changes this behavior, creating a 'source-sink' signature around $\nu \simeq 735\,{\rm MHz}$.}
\label{fig:DT}
\end{figure}

In what follows we allow for redshift evolution of the amplitude but assume that the spectral index of the radio background does not vary with redshift. 
We express the additional radio occupation number as $\Delta n_R(z)=\alpha(z)\,x^{-\gamma}$ with $\alpha(z)=8.84 \,x_0^{\gamma-1}\,f(z)\approx \pot{1.22}{-5}\,f(z)$ and $\gamma=3.59$. Inserting this into Eq.~\eqref{eq:scattering}, we then find
\begin{align}
\left.\frac{\Delta I}{I}\right|_{\rm R}
&\approx y\,\gamma(\gamma-3),
\end{align}
with respect to the initial radio intensity spectrum at the cluster's redshift, $z_{\rm c}$. 
Assuming that $x\ll 1$ (i.e., we are considering observations at radio frequencies), the CMB signal is well into the Rayleigh-Jeans regime (where $\gamma_{\rm CMB} \simeq 1$); we then have 
$\Delta I/I|_{\rm tSZ}\approx \Delta T/T_{\rm CMB}|_{\rm tSZ}\approx-2y$ at $z=0$, independently of the cluster's redshift. For the radio background, we need to more generally write 
\begin{align}
\left.\frac{\Delta T}{T_{\rm R}}\right|_{z=0}
&\approx y\,f(z_{\rm c})\,\gamma(\gamma-3)
\end{align}
to account for the possible redshift dependence of the mean cosmic radio background.
In total we then have the change of the brightness temperature with respect to the unscattered background brightness temperature $T_{\rm R}+T_{\rm CMB}$ as
\begin{align}
\Delta T(z=0)
&\approx y \left[T_{\rm R, 0}(\nu)\,f(z_{\rm c})\,\gamma(\gamma-3) - 2 T_{\rm CMB} \right].
\label{eq:radioSZ}
\end{align}
The fractional increment for the radio SZ signal is comparable to the fractional decrement for the CMB SZ signal, but the radio intensity is much larger at lower frequencies, while the CMB is much brighter at higher frequencies. Inevitably, there will be a null, where there is no net SZ effect; the frequency at which this null occurs depends on the amplitude and spectral index of the radio background at the location of the cluster. From Eq.~\eqref{eq:radioSZ}, one finds that 
\begin{align}
\nu_{\rm R, c}&\approx \nu_0\,\left[\frac{T_{R, 0}(\nu_0)\,f(z_{\rm c})\,\gamma(\gamma-3)}{2 T_{\rm CMB}}\right]^{1/(\gamma-1)}
\label{eq:radioSZ_null}
\end{align}
marks the position of the radio SZ null, which is independent of $y$, just like for the standard thermal SZ effect. With $T_{R, 0}(\nu_0)=24.1\,{\rm K}$ and $\gamma=3.59$, one then finds $\nu_{\rm R, c}\simeq 735 \,{\rm MHz} \,[f(z_{\rm c})]^{0.39}$. At $\nu\gg \nu_{\rm R, c}$, the standard SZ effect arises, with the well-known SZ null around $\nu_{\rm c}\approx 217 \,{\rm GHz}$. On the other hand at $\nu\lesssim \nu_{\rm R, c}$, the scattering effects related to the radio background are more important and the signal increases steeply towards lower frequencies (Fig.~\ref{fig:Intensity} and \ref{fig:DT}).

In the derivation above, the radio background is assumed to be isotropic. For a purely cosmological origin of the radio background, this would be highly justified, but modifications to the expressions given above would arise if not. We furthermore assumed that the up-scattering of radio photons is caused by non-relativistic thermal electrons. Relativistic temperature corrections, just like for the standard SZ effect \citep{Sazonov1998, Itoh98, Chluba2012SZpack}, are expected to become important for typical cluster temperatures ($kT\simeq 5$~KeV), increasing the efficiency of up-scattering at a fixed $y$-parameter and thus shifting the radio null upward. Using estimates for the electron temperature, e.g., observational from X-rays or theoretical from simulations, allows one to include this effect into the analysis once the tSZ for the cluster has been characterized in the standard CMB bands. 

Finally, the equivalent of the kinematic SZ effect \citep{Sunyaev1980} will also arise at radio frequencies. The signal is given by
\begin{align}
\left.\frac{\Delta I}{I}\right|_{\rm R}
&\approx -\tau\,\beta_{\rm p, \parallel}\, x \partial_x \ln \Delta n_{\rm R}= \tau\,\beta_{\rm p, \parallel} \gamma,
\end{align}
where $\beta_{\rm p, \parallel}=v/c$ is the line of sight component of the clusters peculiar motion. For the standard kinematic SZ effect one has $\left.\Delta I/I\right|_{\rm kSZ}=x \expf{x}/(\expf{x}-1)\,\tau\,\beta_{\rm p, \parallel}$, which becomes $\left.\Delta T/T\right|_{\rm kSZ}\approx \tau\,\beta_{\rm p, \parallel}$ at $x\ll 1$. In contrast, the radio kinematic SZ effect is amplified by a factor of $\gamma\simeq 3.59$ due to the spectral steepness of the radio background.

At low frequencies ($\nu\lesssim 10$~GHz), this implies a total signal of 
\begin{align}
\Delta T(z=0)
&\approx y \left[T_{\rm R, 0}(\nu)\,f(z_{\rm c})\,\gamma(\gamma-3) - 2 T_{\rm CMB} \right]+\tau \beta_{\rm p, \parallel} \left[T_{\rm R, 0}(\nu)\,f(z_{\rm c})\,\gamma + T_{\rm CMB}\right].
\label{eq:radioSZ_tot}
\end{align}
when adding the thermal and kinematic SZ effect from the CMB and the radio background. For clusters moving towards the observer, the brightness temperature is increased, as expected. Relativistic kinematic corrections can be obtained similar to the standard kinematic SZ effect \citep{Sazonov1998, Nozawa1998SZ, Challinor2000, Chluba2005b, Chluba2012SZpack}, which would be required for high precision modeling. We note that because the scattering is happening at low-frequencies, where all incoming signals are power-laws, the  relativistic corrections can be obtained from a simple asymptotic expansion of the Compton collision term, without suffering from problems that plague this approach for the standard SZ effects at high frequencies \citep[e.g.,][]{Chluba2012SZpack}.

\section{Detectability of the Signal}
Massive clusters of galaxies can have central Comptonization values  $y>10^{-4}$, which leaves the fractional effects small but possibly still measurable. Detection would require a differential measurement, where the effect at the location of the cluster is compared to a surrounding region without a cluster, or with a matched filter that achieves something similar in an optimal way. Also Map-based constrained ILC methods \citep[e.g.,][]{Remazeilles2011, Remazeilles2021} could be used directly on the radio data to look for source-sink structures in the sky and create a radio $y$-map.

In temperature units, the shift is expected to be a fraction of a mK at frequencies just below 1 GHz. In terms of integration time, a receiver with a system temperature of order 100\, K over a bandwidth $\sim 100$ MHz could reach these levels in just a few hours of integration, if the only concern were instrument noise.

At long radio wavelengths, these sub-mK amplitude shifts are similar to those expected for the $z\sim 1$ redshifted 21cm fluctuations. As a result, experiments that are searching for intermediate-$z$ 21cm fluctuations will have more than sufficient sensitivity to be able to measure this effect. As with the redshifted 21cm fluctuations, the major challenge will be to separate the radio SZ effect from Galactic foregrounds and extragalactic point sources, as well as separating other radio emission from galaxy clusters such as radio halos and radio relics \citep{vanweeren2019}. On the other hand, the frequency dependence is well-predicted, and its spatial dependence precisely follows the $y$ profiles that are well measured at CMB frequencies. Similarly, the kinematic effect can also be measured at CMB frequencies and provide a clear radio spatial and spectral template.

We can convert the expected signal for the thermal effect into flux density:
\begin{equation}
\Delta S(\nu, z_c)
\simeq 0.9\,{\rm mJy} \left[\frac{y \Omega}{10^{-9}}\right] \left[\frac{\nu}{\nu_{\rm R, c}}\right]^2 \left[\frac{T_{\rm R, 0}(\nu)\,f(z_{\rm c})}{2T_{\rm CMB}}\,\gamma(\gamma-3) - 1 \right] 
\simeq 0.9\,{\rm mJy} \left[\frac{y \Omega}{10^{-9}}\right] \left[ \frac{\nu}{\nu_{\rm R, c}}\right]^2 \left[\left(\frac{\nu}{\nu_{\rm R, c}}\right)^{-\kappa} - 1 \right],
\label{eq:radioSZ_flux}
\end{equation}
where $\nu_{\rm R, c}\simeq 735$~MHz is given by Eq.~\eqref{eq:radioSZ_null}, $\kappa=\gamma-1\approx 2.59$ and $\Omega$ is the effective area of the source. A typical value of $y\Omega$ for distant very massive clusters has been assumed \citep{Planck2013SZ, Plagge2010}. This is a small flux, but within the reach of current and planned radio surveys. In terms of raw signal-to-noise, this signal is within the reach of a large number of existing and near-future experiments, including CHIME \citep{chime2021}, ASKAP \citep{askap2021}, MeerKAT \citep{Jonas2016Meerkat}, Tianlei \citep{tianlei2021}, and GBT \citep{gbt2021}. 

A larger concern is the radio emission from the galaxy cluster, including AGN activity and star formation from cluster members and radio halos and radio relics in the intracluster medium. Modeling this cluster radio emission at the same time as the radio SZ effect will be challenging but not necessarily impossible. The SZ signal has a distinct spectral shape (for a given radio background amplitude), with an increment at low frequencies and a decrement at high frequencies, and the $y$ parameter for the cluster is likely to be known from mm-wave observations, giving a precise template for the signal of interest. This signal will be different from the expected smooth power-law behavior of the radio synchrotron emission from AGN, star formation, and radio halos and relics.

A possible approach would be to not focus on the spectral window around the ``radio null'' and instead simply look for a deviation at low frequencies from the well-studied regular SZ effect in the Rayleigh-Jeans tail of the CMB. Rather than the standard "$-2y$'' compared to the CMB, at GHz frequencies there should be noticeable weakening of the SZ signal in clusters at $\nu\lesssim 3$~GHz (see Fig.~\ref{fig:DT}). While the deviation is smaller at frequencies of a few GHz, the foreground issues could be reduced, just as early detection of the thermal SZ effect were not done near the null at 217~GHz but were instead done at radio frequencies \citep{birkinshaw1978}.

Furthermore, one could simply use maps of the Compton $y$-parameter from Planck, ACT, and/or SPT as a spatial template for a cross-correlation analysis \citep[e.g.,][]{Planck2013ymap,actymap2020,bleem2021}. Assuming the radio background is in place before most of the Comptonization, the radio SZ map will be simply a scaled copy of the CMB-derived $y$-maps, with the frequency-dependent amplitude set by the power-law index of the radio background and the radio SZ amplitude directly determined by the amplitude of the radio background. Similarly, avenues of cluster stacking analysis can be considered.

\section{Discussion \& Conclusion}

We have shown that a radio background leads to a clear Comptonization signal at low frequencies, with a null that is determined by the amplitude of the radio background. For the radio background inferred by ARCADE and others, this radio SZ signal is faint but in principle detectable with current instruments.

Detection of this radio SZ effect would be a direct confirmation that the measured radio background really is as high as suggested by measurements, despite the difficult theoretical interpretation.
Measuring this signal as a function of redshift will measure the radio background as a function of redshift, perhaps giving clues to the origin of the signal.

The predicted radio SZ has a steep scaling towards low frequencies (see Fig.~\ref{fig:DT}). As such, it could become an important foreground for standard 21cm fluctuation measurements. The signal is expected to spatially correlate with the standard Compton-$y$ map and therefore can be modelled by combining SZ and 21cm measurements. This is of particular interest to radio experiments such as LOFAR \citep{Haarlem2013} and the Square Kilometer Array  \citep{Bacon2020}, and CMB experiments like the Simons Observatory \citep{SOWP2018} and CMB-S4 \citep{cmbs4_2019}, which all promise high resolution wide-area maps of the sky.

The corresponding kinematic effect will be less contaminated by radio emission from galaxy clusters, as it gets contribution equally from all ionized gas, while the thermal effect is largest in the deep potential wells of galaxy clusters. This will allow high-redshift tomography of the radio background, tracing large scale structure out to the epoch of reionization. Due to the distinct spectral signature, this probe is furthermore not degenerate with the primordial CMB fluctuations that hamper the constraining power of the standard kinematic SZ.

More broadly speaking, our calculations highlight the importance of modifications to the SZ effect caused by additional nearly-isotropic contributions at the cluster location. These will up-scatter and contaminate the SZ signal template in potentially non-trivial ways. For future high precision cosmological applications of SZ cluster samples, this could become of importance. Scattering of the ambient cosmic infrared background is one clear concern, in particular at high frequencies ($\nu\gtrsim 400$~GHz). In addition, scattering of the cosmic CO emission could become noticeable. 

These effects could evidently be problematic, e.g., when trying to extract the relativistic SZ \citep{Hansen2002, Erler2017, Hurier2017rSZ, Remazeilles2020} or using the SZ effect to measure the Hubble constant \citep{Hughes1998, Bonamente2006},
study the redshift-dependence of the CMB temperature \citep{Rephaeli1980, Luzzi2009, Li2021} or constrain variations of fundamental constants \citep{Galli2013, Holanda2019}. However, this also provides the unique opportunity to use clusters as a probe of the ambient light in the distant universe. Similar to the scattering of the local CMB quadrupole by SZ clusters \citep[e.g.,][]{Kamionkowski1997b, Sazonov2000}, one can also probe the local radio and CIB quadrupole, further enriching the range of possibilities. Finally, by studying the effect in different directions of the sky one can probe the isotropy of the additional background contributions.

\small
\vspace{3mm}
{\it Acknowlegdement}: The authors cordially thank the Aspen Center for Physics, which is supported by National Science Foundation grant PHY-1607611, for their great hospitality in September 2021, where this work was initiated. As part of the the visit to the Aspen Center for Physics, this work was supported by the Simons Foundation.
GH received additional support from CIfAR and Brand \& Monica Fortner.
JC thanks Keith Grainge, Ralph Spencer and Benjamin Stappers for useful discussions about radio observations. We are also very grateful for comments from Mathieu Remazeilles, Jack Singal, Kimmy Wu and Neelima Sehgal. 
JC was furthermore supported by the ERC Consolidator Grant {\it CMBSPEC} (No.~725456) and the Royal Society as a Royal Society University Research Fellow (URF/R/191023).

\bibliographystyle{aasjournal}
\bibliography{Lit}

\end{document}